# A nonparametric control chart based on the Mann-Whitney statistic

## Subhabrata Chakraborti[1] and Mark A. van de Wiel[2]

*University of Alabama and Vrije Universiteit Amsterdam*

**Abstract:** Nonparametric or distribution-free charts can be useful in statistical process control problems when there is limited or lack of knowledge about the underlying process distribution. In this paper, a phase II Shewhart-type chart is considered for location, based on reference data from phase I analysis and the well-known Mann-Whitney statistic. Control limits are computed using Lugannani-Rice-saddlepoint, Edgeworth, and other approximations along with Monte Carlo estimation. The derivations take account of estimation and the dependence from the use of a reference sample. An illustrative numerical example is presented. The in-control performance of the proposed chart is shown to be much superior to the classical Shewhart $\bar{X}$ chart. Further comparisons on the basis of some percentiles of the out-of-control conditional run length distribution and the unconditional out-of-control $ARL$ show that the proposed chart is almost as good as the Shewhart $\bar{X}$ chart for the normal distribution, but is more powerful for a heavy-tailed distribution such as the Laplace, or for a skewed distribution such as the Gamma. Interactive software, enabling a complete implementation of the chart, is made available on a website.

## 1. Introduction

Control charts are most widely used in statistical process control (SPC) to detect changes in a production process. In conventional SPC, the pattern of chance causes is often assumed to follow the normal distribution. It is well recognized however that in many applications the underlying process distribution is not known sufficiently to assume normality (or any other parametric distribution), so that statistical properties of commonly used charts, designed to perform best under the assumed distribution (such as normality), could be potentially (highly) affected. In situations like this, development and application of control charts that do not depend on normality, or more generally, on any specific parametric distributional assumptions, seem highly desirable. Distribution-free or nonparametric control charts can serve this purpose. Chakraborti, Van der Laan and Bakir [4] (hereafter CVB) presented an extensive overview of the literature on univariate nonparametric control charts. In-control (stable) properties of these charts are completely determined (known)

*Supported in part by a summer research grant from the University of Alabama and by a grant from the Thomas Stieltjes Institute for Mathematics.
[1]Department of Information Systems, Statistics and Management Science, University of Alabama, Tuscaloosa, Alabama, USA, e-mail: schakrab@cba.ua.edu
[2]Department of Mathematics, Vrije Universiteit Amsterdam, Amsterdam, The Netherlands, e-mail: mark.vdwiel@vumc.nl

and remain the same for all continuous distributions and hence their out-of-control behavior are more meaningful and comparable.

In this paper we use the well-known Mann-Whitney test statistic as a charting statistic for detecting location shifts. The problem of monitoring the center or the location of a process is important in many applications. The location parameter could be the mean or the median or some percentile of the distribution; the latter two are often more attractive when the underlying process distribution is expected to be skewed. Among the available control charts for the mean of a process, the classical Shewhart $\bar{X}$ chart is the most popular because of its inherent simplicity and practical appeal. In some applications, the process distribution and/or the parameters are specified or can be assumed known. This is typically referred to as the standards known case (Montgomery [11], page 228). If on the other hand the parameters are unknown and are estimated from data, there is growing evidence in the recent literature that most standard charts, including the Shewhart $\bar{X}$ chart, behave quite poorly in terms of the false alarm rate and the average run length. We do not assume that the process parameters are specified or that the process distribution is known, instead, we assume that a reference sample is available from the in-control process from a phase I analysis. Once the control limits are determined from the reference sample, monitoring of test samples is begun. This is referred to as a phase II application.

There are some phase II nonparametric charts available in the literature; the reader is referred to CVB for many references and detailed accounts. The nonparametric charts considered by Chakraborti, Van der Laan and Van de Wiel [5] (hereafter CVV) are based on the precedence test. It is seen that the precedence charts are good alternatives to the $\bar{X}$ chart in some situations. However, while the precedence charts are a step in the right direction, it is known that the nonparametric test underlying this chart, the precedence test, is neither the most powerful test (for location) nor the most commonly used nonparametric test in practice. With this motivation, we consider a chart based on the popular and more powerful Mann-Whitney [9] (hereafter MW) test which is equivalent to the perhaps more familiar Wilcoxon [15] rank-sum test. This is called the MW control chart.

One might suspect that the distribution-free-ness of the MW chart might come at a "loss of power" with respect to parametric charts. However, remarkably, even when the underlying distributions are normal, the MW test is about 96% as efficient (Gibbons and Chakraborti [7], pages 278–279) as (the most efficient) t-test for moderately large sample sizes, and yet, unlike the t-test, it does not require normality to be valid. Park and Reynolds [12] realized the potential of nonparametric control charting and introduced a chart based on this statistic. They considered various properties of this chart when the reference sample size approaches infinity, which essentially amounts to assuming the standards known case. While this is important for theoretical purposes and gives some insight, such a chart does not appear to be very useful in practice where parameters and/or the underlying distribution are unknown and need to be estimated from a moderate size phase I data set. In fact, it is crucial to develop and implement the MW chart in practice for small to moderate reference sample sizes since, as we show later in the paper, the MW chart can be especially useful in such cases.

While the principles of this chart are simple, practical implementation of the chart, i.e., developing an efficient algorithm for computing the control limits, requires some effort. We provide software for practical use of the chart. Effectiveness of the chart is examined on the basis of several in-control and out-of-control performance criteria. We conclude with a discussion, including some topics for further research.

## 2. The MW control chart

Suppose that a reference sample of size $m$, denoted by $X = (X_1, \ldots, X_m)$, is available from an in-control process and that $Y = (Y_1, \ldots, Y_n)$ denotes an arbitrary test sample of size $n$. The superscript $h$ is used to denote the $h^{th}$ test sample, $Y^h = (Y_1^h, \ldots, Y_n^h)$, $h = 1, 2, \ldots$, when necessary for notational clarity; otherwise, the superscript is suppressed. Assume that the test samples are independent of each other and are all independent of the reference sample. The MW test is based on the total number of $(X, Y)$ pairs where the $Y$ observation is larger than the $X$. This is the statistic

$$(2.1) \quad M_{XY} = \sum_{i=1}^{m}\sum_{j=1}^{n} I(X_i < Y_j) = \sum_{j=1}^{n}\{I(Y_j > X_1) + \cdots + I(Y_j > X_m)\},$$

where $I(X_i < Y_j)$ is the indicator function for the event $\{X_i < Y_j\}$. Note that $M_{XY}$ lies (attains values) between $0$ and $mn$ and large values of $M_{XY}$ indicate a positive shift, whereas small values indicate a negative shift.

The proposed two-sided MW chart uses $M_{XY}^h$ as the charting statistic, which is $M_{XY}$ for the $h^{th}$ test sample. The chart signals if

$$M_{XY}^h < L_{mn} \text{ or } M_{XY}^h > U_{mn},$$

where $L_{mn}$ and $U_{mn}$ are the lower control limit (LCL) and the upper control limit (UCL), respectively. The distribution of $M_{XY}$ is known to be symmetric about $mn/2$ when the process is in-control, so it is reasonable to take $L_{mn} = mn - U_{mn}$. We focus on two-sided charts, one-sided charts can be developed similarly.

### 2.1. Design and implementation

Implementation of the chart requires the control limits. Typically, in practice, the control limits are determined for some specified in-control average run length ($ARL_0$) value, say 370 or 500. If the successive charting statistics $M_{XY}^1, M_{XY}^2, \ldots$ corresponding to test sample $1, 2, \ldots$ were independent, then, as in the standards known case, the $ARL_0$ would be equal to the reciprocal of the false alarm rate, $p_0 = 2P_0(M_{XY} > U_{mn})$, where the subscript 0 denotes the in-control case. So if the charting statistics were independent, the upper control limit $U_{mn}$ would be equal to the two-sided critical value for a MW test with size equal to $1/ARL_0$. However, such critical values are not expected to be found in available tables for the MW test, since in a typical control charting application $ARL_0 = 370$, which means that the $UCL$ is the upper critical value for a MW test with size $\frac{1}{2}(1/370) = 0.00135$.

Even if such critical values could be found, the main problem with their use is that the successive charting statistics $M_{XY}^1, M_{XY}^2 \ldots$ are dependent, since the test samples are all compared to the same control limits derived from the same reference sample, and this dependence affects all operational and performance characteristics of the control chart, such as the false alarm rate, the $ARL$, etc. (see, for example, Quesenberry [13] and Chakraborti [3] for the Shewhart $\bar{X}$ chart). It might be argued that for "large" amounts of reference data such dependence can be ignored. There are two problems with this argument, however. One, we would need to know the size of the reference data, a-priori, that would support ignoring dependence, and two, we would have to wait much longer to gather that amount of data while process monitoring has to wait, costing time and money. The solution to this is the ability to calculate the control limits for any given (fixed) $m$ and $n$ and $ARL_0$ in a given situation. To this end we first develop an efficient method to calculate the $ARL$.

## 2.2. Calculation of the ARL

Let $F$ and $G$ represent the cdf of $X$ and $Y$ respectively, and suppose that $F$ and $G$ are continuous, so that "ties" between the $X$'s and the $Y$'s, as well as within the $X$'s and the $Y$'s can be ignored theoretically. It is convenient to derive the $ARL$ by conditioning on the reference sample, i.e. using the so-called conditioning method. To this end, observe that the probability of signal for any test sample, given the reference sample $X = x$, is

$$(2.2) \qquad p_G(x) = P_G(M_{xY} < mn - U_{mn}) + P_G(M_{xY} > U_{mn}).$$

Let $N$ denote the run length random variable for the chart. Given the reference sample $X = x$, and that two arbitrary test samples $Y^h$ and $Y^l$, $(h \neq l)$ are independent, which implies the independence of $M_{xY}^h$ and $M_{xY}^l$. Hence,

$$(2.3) \begin{aligned} ARL = E(N) &= E_F[E_G(N|X=x)] = E_F(\frac{1}{p_G(x)}) \\ &= \int_{-\infty}^{\infty} \cdots \int_{-\infty}^{\infty} \frac{1}{p_G(x)} \, dF(x_1) \cdots dF(x_m) \\ &= \int_{-\infty}^{\infty} \cdots \int_{-\infty}^{\infty} \nu(G(x_1), \ldots, G(x_m)) \, dF(x_1) \cdots dF(x_m), \end{aligned}$$

say. The second equality in (2.3) follows from a property of expectation. The third equality follows since given $X = x$, $N$ is geometrically distributed with parameter $p_G(x)$. The fourth equality is obtained by writing $1/p_G(x)$ as a function of $G$ and $x_1, \ldots, x_m$, say, $\nu(G(x_1), \ldots, G(x_m)) = \nu(P_G(Y_j < x_1), \ldots, P_G(Y_j < x_m))$, where $\nu$ is some function. This can be done since $p_G(x)$ is a sum of probabilities like $P_G(M_{xY} = u)$, which, in turn is a sum of probabilities over all configurations of the $x$'s and $Y$'s for which $M_{xY}$ equals $u$. Naturally, the probability of such a configuration only depends on $G(x_1), \ldots, G(x_m)$.

The (unconditional) $ARL$ of the chart is the mean (expectation) of the distribution of the random variable $E_G(N|X)$, which is the conditional average run length, given the random reference sample. Percentiles of the distribution of $E_G(N|X)$ (and not just the mean) are useful to study and characterize control chart performance when parameters are estimated, and we develop efficient algorithms to compute these. First however, we focus on the mean of the conditional distribution, that is the unconditional $ARL$ given in (2.3), when the process is in-control.

In the in-control situation the $X$'s and the $Y$'s come from the same distribution

$F = G$. Therefore, we may assume w.l.o.g. that $F = U[0, 1]$:

$$
\begin{aligned}
ARL_0 &= \int_{-\infty}^{\infty} \cdots \int_{-\infty}^{\infty} \nu(F(x_1), \ldots, F(x_m))\, dF(x_1) \cdots dF(x_m) \\
&= \int_0^1 \cdots \int_0^1 \nu(u_1, \ldots, u_m)\, du_1 \cdots du_m \\
&= \int_0^1 \cdots \int_0^1 \frac{1}{p_U(u)}\, du_1 \cdots du_m,
\end{aligned}
\tag{2.4}
$$

where $p_U(u) = P_U(M_{uY} < mn - U_{mn}) + P_U(M_{uY} > U_{mn})$ is the conditional probability of a signal at any test sample, given the reference sample $u$, when the process is in-control. The subscript $U$ is used to denote that in the in-control case both the reference and the test samples can be thought of coming from the same distribution, the $U[0,1]$ distribution, which shows that the in-control $ARL$ of the proposed chart does not depend on $F$. The same argument can be used to show that the in-control run-length distribution does not depend on $F$ and hence the proposed chart is distribution-free. We emphasize that for the in-control case, without any loss of generality, the common ($F = G$) distribution can be assumed to be the $U[0,1]$ distribution by virtue of the probability integral transform (see, for example, Gibbons and Chakraborti [7]). This simplifies calculations considerably.

We need to calculate (2.4) to implement the chart and (2.3) to evaluate chart performance. For both of these objectives there are two problems. First, an explicit formula for $p_G(x)$ (or $p_U(u)$), is not known, which prevents a direct computation. Second, we have an $m$-dimensional integration in both (2.3) and (2.4). Our approach is to calculate $p_G(x)$ (and $p_U(u)$) (exactly or approximately) using a fast algorithm, and then use that to approximate the integral in both (2.3) and (2.4) with a Monte Carlo estimate to get estimates

$$
\hat{ARL}_G \approx \frac{1}{K} \sum_{i=1}^{K} \frac{1}{p_G(x_i)}
\tag{2.5}
$$

and

$$
\hat{ARL}_0 \approx \frac{1}{K} \sum_{i=1}^{K} \frac{1}{p_U(u_i)},
\tag{2.6}
$$

where $x_{i\cdot} = (x_{i1}, \ldots, x_{im})$ is the $i^{th}$ Monte Carlo sample, $i = 1, \ldots, K$, of which each component is drawn from some specified $F$ for the $\hat{ARL}_G$, and $K$ denotes the number of Monte Carlo samples used. Similarly, for the in-control situation, $u_i$ is a Monte Carlo sample from $U[0,1]$. For an accurate approximation, $K$ needs to be sufficiently large and therefore a fast method of computing the signal probability $p_G(x)$ (and $p_U(u)$) for an arbitrary reference sample $x$ is essential for the practical use of the approximation.

The $ARL$ calculations proceed in two steps. The first step is to find a fast and efficient method to compute the signal probability. We detail the procedure for fast computation of $p_G(x)$; calculation of $p_U(u)$ is similar.

## *2.3. Fast computation of signal probability*

From (2.1) it is seen that the MW statistic is the sum $\sum_{j=1}^{n} C_j$, where $C_j$ represents the number of $X$'s that are less than an $Y_j$. From (2.2) it follows that the calculation of $p_G(x)$ essentially requires calculation of the upper-tail probability $P_G(M_{xY} > U_{mn})$, say, and this in turn requires (i) efficient enumeration of all $n$-tuples $\{C_1, \ldots, C_n\}$ for which the sum is greater than $U_{mn}$ and (ii) summation of the probabilities for such tuples.

Note that $P(C_j = l)$ is equal to $P(X_{(l)} < Y_j < X_{(l+1)})$, where $X_{(l)}$ denotes the $l^{th}$ ordered observation in the reference sample for $l = 1, \ldots, m$ with $X_{(0)} = -\infty$ and $X_{(m+1)} = \infty$, say. Given the reference sample, the last probability is simply $P(x_{(l)} < Y_j < x_{(l+1)})$, which is denoted by $a_l$, $l = 0, \ldots, m$. Also, given the reference sample, note that the random variables $C_j$ are i.i.d. Hence the conditional probability generating function (pgf) of $C_j$ is

$$(2.7) \qquad H_1(z) = \sum_{l=0}^{m} P(C_j = l) z^l = \sum_{l=0}^{m} a_l z^l.$$

Again, since given the reference sample the $C_j$ are i.i.d., the conditional pgf of $M_{xY}$ (the sum of the $C_j$), is simply the product of the pgf's in (2.7)

$$(2.8) \qquad H_2(z) = \sum_{j=0}^{mn} P(M_{xY} = j) z^j = (\sum_{j=0}^{m} a_j z^j)^n.$$

In principle, $P_G(M_{xY} > U_{mn})$ can be calculated by expanding the power in (2.8) and collecting the coefficients of all terms with degree greater than $U_{mn}$. However, for moderate to large $m$ (say, $m \geq 100$) and $n$ not very small (say, $n \geq 5$) this takes a considerable amount of computing time, especially since the procedure has to be repeated $K$ (a large number) times, once for each Monte Carlo sample. Alternative, faster, methods such as "a branch-and-bound" algorithm, based on Mehta et al. [10] can be used to shorten the intermediate expressions that result from expanding (2.8) and saves considerable time.

However, even with the branch-and-bound algorithm, $m$ and $n$ might be just too large to allow for exact computations and hence a good approximation to the control limits may be necessary. For example, we may apply the central limit theorem for the sum of i.i.d. random variables to $M_{xY} = \sum_{j=1}^{n} C_j$ to get a normal approximation to $P_G(M_{xY} > U_{mn})$ but in our context, $U_{mn}$ is typically far in the upper tail of the distribution of $M_{xY}$ and $n$ is usually not very large, and so the normal approximation is not very accurate. Instead, we find the Lugannani-Rice formula (hereafter LR-formula; see Jensen [8], page 74) for the upper-tail probability for the mean of i.i.d. discrete random variables (which is a "saddlepoint" approximation formula) to be particularly useful. This formula is known to be more accurate than the normal approximation in the tails of a distribution and is based on the cumulant generating function of $C_j$, which is obtained from the pgf in (2.7): $k(t) = \log[H_1(e^t)]$. Let $m(t)$ and $\sigma^2(t)$ denote the first and the second derivative of $k(t)$, respectively. Furthermore, let $u = (U_{mn} + 1)/n$ and $\bar{M}_{xY} = M_{xY}/n$. The saddlepoint $\gamma$ is the solution to the equation $m(t) = u$. Using (3.3.17) in Jensen (1995, page 79) we obtain

$$(2.9) \qquad \begin{aligned} P_G(M_{xY} > U_{mn}) &= P_G(\bar{M}_{xY} > U_{mn}/n) = P_G(\bar{M}_{xY} \geq u) \\ &\approx 1 - \Phi(r) + \phi(r)(\frac{1}{\lambda} - \frac{1}{r}), \end{aligned}$$

where
$$\lambda = n^{1/2}(1 - e^\gamma)\sigma(\gamma), \ r = (sgn\gamma)\{2n(\gamma u - k(\gamma))\}^{1/2},$$

$\Phi(.)$ and $\phi(.)$ are, respectively, the cdf and the pdf of the standard normal distribution. Using (2.9), we can efficiently approximate the signal probability $p_G(x)$ given in (2.2).

## 2.4. Monte Carlo estimation of ARL and error control

From formulas (2.5) and (2.6) we observe that the computation of $p_G(x)$ is repeated many times to obtain a Monte Carlo approximation of the $ARL$. The question here is regarding $K$, the number of samples so that the Monte Carlo error is acceptably small. Since, for the purpose of $ARL$ computation, the reference samples are drawn independently from $G$ (or $U[0,1]$ in case of $ARL_0$), the Monte Carlo standard error is estimated by $s_{mc} = s(ARL_G(X))/\sqrt{K}$, where $s(.)$ denotes the sample standard deviation computed from $K$ simulated reference samples. Then, we may choose the smallest $K$ such that

$$(2.10) \qquad s_{mc} = s(ARL_G(X))\big/\sqrt{K} \leq D,$$

where $D$ is either a specified number or a percentage of the current estimate $\hat{ARL}_G$. We can start with say $K = 100$, increase $K$, compute $s_{mc}$, and repeat the process until the specification $s_{mc} \leq D$ is met. Use of formula (2.10) provides a way to obtain an efficient and a reasonably accurate approximation to $ARL_G$. Asymptotic probabilistic control of the accuracy may be achieved by using the normal distribution for $\hat{ARL}_G$. For example, one could set $D$ such that the probability that $\hat{ARL}_G$ deviates more than $C$ units from the real mean is smaller than 0.05. Here, $C$ could be a small percentage of the current estimate. Next, we discuss the approximation of $ARL_0$ in more detail.

*Approximation of $ARL_0$*

Three methods of approximating $ARL_0$ have been introduced so far, each based on a different method to compute or approximate $p_U(u)$. To summarize, they are:

1. Exact (EX): Monte Carlo simulation using (2.6), with $p_U(x)$ computed exactly using formula (2.8)
2. LR-formula (LR): Monte Carlo simulation using (2.6), with $p_U(u)$ computed approximately using formula (2.9)
3. Normal (NO): Monte Carlo simulation using (2.6), with $p_U(u)$ computed from a normal approximation

We compare these methods on the basis of accuracy and speed. While setting a value of $D$, we observed that $K$ was usually under 1000, and the maximum Monte Carlo error was 2% of the target $ARL_0 = 500$, that is, equal to 10. The value of $K$ was set to 1000 and kept unchanged in these computations to get a fair comparison of the computing times. Moreover, we introduce two alternative methods to calculate $ARL_0$:

4. Fixed reference sample (FR): Fix reference sample to $q = (1/(m+1), \ldots, m/(m+1))$ and approximate $ARL_0$ by $1/p_U(q)$
5. 1/(false alarm rate) (FA): approximate $ARL_0$ by the reciprocal of the false alarm rate: $1/(2P_0(M_{XY} > U_{mn}))$.

TABLE 1
$ARL_0$ approximations and computing times

| | | EX | | LR | | NO | | FR | | FA | |
|---|---|---|---|---|---|---|---|---|---|---|---|
| | | $A\hat{R}L_0$ | time (sec.) | $A\hat{R}L_0$ | time (sec.) | $A\hat{R}L_0$ | time (sec.) | $A\hat{R}L_0$ | time (sec.) | $A\hat{R}L_0$ | Time (sec.) |
| $m$ | $n$ | | | | | | | | | | |
| 50 | 5 | 486 | 54 | 506 | 36 | 307 | 1.0 | 403 | 0.05 | 247 | 0.01 |
| | 10 | 504 | 395 | 505 | 34 | 327 | 1.0 | 524 | 0.05 | 226 | 0.01 |
| | 25 | 488 | 4850 | 491 | 31 | 425 | 1.2 | 694 | 0.05 | 119 | 0.01 |
| 100 | 5 | 496 | 220 | 505 | 48 | 219 | 1.2 | 478 | 0.05 | 353 | 0.01 |
| | 10 | 505 | 1920 | 506 | 47 | 339 | 1.3 | 531 | 0.05 | 332 | 0.01 |
| | 25 | ** | 26168 | 503 | 48 | 422 | 1.3 | 683 | 0.06 | 233 | 0.01 |
| 500 | 5 | 491 | 10633 | 496 | 207 | 226 | 1.2 | 492 | 0.20 | 445 | 0.01 |
| | 10 | ** | 73516 | 513 | 179 | 367 | 1.7 | 537 | 0.21 | 484 | 0.01 |
| | 25 | ** | $7.59*10^5$ | 494 | 176 | 445 | 1.6 | 578 | 0.29 | 450 | 0.01 |
| 1000 | 5 | ** | 31766 | 500 | 356 | 235 | 2.1 | 513 | 0.48 | 471 | 0.01 |
| | 10 | ** | $3.42*10^5$ | 499 | 373 | 355 | 2.4 | 516 | 0.49 | 488 | 0.01 |
| | 25 | ** | $3.15*10^6$ | 500 | 348 | 442 | 1.7 | 548 | 0.63 | 482 | 0.01 |
| 2000 | 5 | ** | $1.71*10^5$ | 503 | 713 | 234 | 2.1 | 506 | 0.67 | 474 | 0.01 |
| | 10 | ** | $1.44*10^6$ | 504 | 659 | 354 | 1.9 | 513 | 0.71 | 499 | 0.01 |
| | 25 | ** | $1.29*10^7$ | 509 | 676 | 446 | 2.1 | 531 | 1.41 | 497 | 0.01 |

\*\* $ARL_0$ could not reliably be estimated within reasonable time; computing time for $K = 1000$ is obtained by multiplying computing time for K = 1 by 1000 (sampling algorithm is linear in K).

Let us explain approximations 4 and 5. When $m$ is large, the empirical cdf $F_m(x)$ converges to $F(x)$ (which is the cdf of the $U[0, 1]$ distribution) and hence for large $m$ we may approximate the $i^{th}$ reference sample observation by the $i/(m+1)^{th}$ quantile of the $U[0, 1]$ distribution, $q_i = i/(m + 1)$, $i = 1, \ldots, m$. Thus, we approximate the $ARL_0$, for large $m$, by $1/p_U(q)$, where $q = (q_1, \ldots, q_m)$. This is method 4. The major benefit with this approximation is that we need to compute $p_U(u)$ only once (namely at $u = q$) instead of $K$ times as needed in methods 1 through 3 for each of $K$ Monte Carlo reference samples. Finally, another quick approximation for the $ARL_0$ is given by the inverse of the false alarm rate, $1/(2P_0(M_{XY} > U_{mn}))$. In a setting where runs are truncated at a finite point $T$, the latter approximation was proven to be unbiased when $m$ approaches infinity in Park and Reynolds [12]. This approximation is method 5. Chakraborti [3] showed that for the Shewhart $\bar{X}$ chart, 1/FAR is a lower bound to $ARL_0$ and noted that this bound can serve as a "quick and dirty" approximation to the $ARL_0$ for moderate to large values of $m$. When applying method 5, we used formula (11) in Fix and Hodges [6] to compute $P_0(M_{XY} > U_{mn})$, based on an Edgeworth approximation, which significantly improves the normal approximation by including moments of order higher than 2.

Table 1 displays the estimated $ARL_0$ values computed by the five methods for fifteen combinations of $m$ and $n$. Chart constants were determined (using the algorithm discussed in the next section) such that $ARL_0 \approx 500$, when applying the exact formula (2.8) or the best approximation, the LR-formula, when exact computations were too time-consuming. Therefore, the closer an $ARL_0$ value is to 500, the better the approximation. The table also shows the computing times on a 1.7GHz Pentium PC with 128MB of internal RAM.

Several observations can be made from Table 1. First we see that the "gold standards" or the exact computations are very time-consuming for most values of $m$ and $n$. However, when they can be found, they would naturally form the basis of our comparisons of various approximations. Second we see that the normal approximation is not very accurate, but the LR-approximation is, particularly for $m \leq 100$, $n \leq 10$ and for $m = 500$, $n = 5$. Since the LR-approximation is known to become more accurate when the sample sizes increase, one may safely apply the LR-formula also when $m \geq 50$ and $n \geq 5$ in order to implement the proposed chart. It may be noted that when $m$ increases, the computing times with the LR–formula also increase, although, by far, not as dramatically as the times for the exact computations. This suggests that in practice (for finding the chart constants, to be discussed next) there is still room for an alternative, quick approximation of $ARL_0$, particularly for large values of $m$. Compared to the LR-formula, we observe that both the "fast" approximations (methods 4 and 5) are quite good for $m \geq 1000$ and that the fixed reference sample approximation (method 4) performs somewhat better for relatively small values of $n$ ($n = 5,10$) than for $n = 25$.

To summarize, the best method of calculating the $ARL_0$ is the exact EX method if it is computationally feasible, otherwise, the best approximation is the LR method. In practice, we recommend using the LR-method, because it is both fast and accurate. If the reference sample is very large, say $m \geq 1000$, one of the two faster approximations, either FR or FA, can be used.

## *2.5. Determination of chart constants*

Since we can now calculate the $ARL_0$ corresponding to a given value of $U_{mn}$ efficiently and accurately, we can use an iterative procedure based on linear interpolation to find the control limit for a pre-specified $ARL_0$ value, say 500. In principle, we use the LR-approximation for the computation of $ARL_0$. However, we have observed that this approximation is still somewhat time-consuming, so we want to minimize the number of iterations for which the LR approximation is used. To this end a good starting value of $U_{mn}$ is needed and this is where the fast approximations FR and FA are very useful. Starting with the FA approximation, we simply equate the inverse of the false alarm rate to 500, which means solving $1/(2 * FH(u)) = 500$ for $u$ in order to get an initial guess for $U_{mn}$. Since the FR approximation is somewhat better than FA for $m \leq 500$, we use this initial guess with the FR approximation to refine the Fix-Hodges approximation when $m \leq 500$. The resulting approximation for the *UCL* is then used as an initial guess for the linear interpolation method with the LR-formula. We do not detail the search procedure here, but illustrate it with an example. Suppose that $m = 375$ and $n = 7$, and we want to find the chart constants such that $ARL_0 \approx 400$. Suppose we allow a deviation of 2% (which is 0.02 * 400 = 8) maximally, hence the search procedure would stop and yield the desired control limit $U_{mn}$, when $392 \leq A\hat{R}L_0 \leq 408$. Moreover, suppose we stipulate that the Monte Carlo standard error $s_{mc}$ be smaller than 1.5% of the current estimate of $A\hat{R}L_0$; from inequality (2.10) we observe that this requirement determines the number of Monte Carlo samples $K$ per iteration, when setting $D= 0.015 * 400 = 6$. The output from our program (written in Mathematica; see Software section later) is shown in Table 2. As can be seen in Table 2, six iterations (numbered 1 through 6) have been executed in approximately 140 seconds. The first three of these hardly take any computing time, because the fast FA and FR approximations were used. For each iteration, the values of the *UCL* and the *LCL* and the corresponding $ARL_0$ are shown. Under the LR method, the program also calculates the $5^{th}$ percentile of the conditional in-control *ARL* distribution and the Monte Carlo standard error ($smc$). Note that the first step with the LR method, step 4, uses the chart constants of step 3, which is our best guess from the fast approximations, as initial values. Also, note that for the LR method, the first two iterations (4 and 5) produce $ARL_0$ values below and above the target value 400, so that linear interpolation begins at the third iteration and the new *UCL* is found using the two previous *UCL*'s

TABLE 2
*Finding control limits for $m = 375$, $n = 7$ and target $ARL_0 = 400$*

**FA: 1/(false alarm rate) approximation**
1. ucl=2146 lcl=479 ARL0=400
**FR: Fixed reference sample approximation**
2. ucl=2146 lcl=479 ARL0=446.761
3. ucl=2136 lcl=489 ARL0=386.729
**LR approximation**
4. ucl=2136 lcl=489 ARL0=380.059 smc=5.69018 5% perc=238.407 K=402
5. ucl=2146 lcl=479 ARL0=438.11 smc=6.5647 5% perc=287.53 K=319
6. ucl=2139 lcl=486 ARL0=394.496 smc=5.91419 5% perc=252.778 K=315
**139.962 Second**

TABLE 3
*Lower and upper MW control chart limits for selected values of $m$, $n$ and $ARL_0$*

| $m$ | $n$ | $ARL_0 = 370$ | | $ARL_0 = 500$ | |
|---|---|---|---|---|---|
| | | $L_{mn}$ | $U_{mn}$ | $L_{mn}$ | $U_{mn}$ |
| 50 | 5 | 35 | 215 | 33 | 217 |
| | 10 | 115 | 385 | 111 | 389 |
| | 25 | 400 | 850 | 393 | 857 |
| 100 | 5 | 69 | 431 | 65 | 435 |
| | 10 | 231 | 769 | 224 | 776 |
| | 25 | 805 | 1695 | 793 | 1707 |
| 500 | 5 | 348 | 2152 | 328 | 2172 |
| | 10 | 1170 | 3830 | 1128 | 3872 |
| | 25 | 4081 | 8419 | 4016 | 8484 |
| 1000 | 5 | 698 | 4302 | 653 | 4347 |
| | 10 | 2344 | 7656 | 2268 | 7732 |
| | 25 | 8169 | 16831 | 8058 | 16942 |
| 2000 | 5 | 1397 | 8603 | 1309 | 8691 |
| | 10 | 4682 | 15318 | 4540 | 15460 |
| | 25 | 16392 | 33608 | 16145 | 33855 |

and their corresponding $ARL_0$ values. Thus, the final chart constants are found at iteration 6, $U_{mn} = 2139$ and hence $L_{mn} = 486$, with attained $ARL_0 = 394.5$. The $5^{th}$ percentile of the conditional in-control $ARL$ distribution, at this iteration, is found to be 252.78. So using the MW chart with $UCL = 2139$ and $LCL = 486$, the unconditional $ARL_0 = 394.5$ implies that when the process is in-control, on an average, a false alarm is expected every 395 samples. The $5^{th}$ conditional percentile implies that for 95% of all reference samples (that could have possibly been taken from the in-control process), the average run length is at least 253.

Table 3 shows chart constants computed with this algorithm for a number of combinations of $m$ and $n$. All cases are two-sided and $ARL_0$ equals either 370 or 500.

## 2.6. Numerical example

Table 5.1 in Montgomery [11] gives a set of data on the inside diameters of piston rings manufactured by a forging process. Twenty-five samples, each of size five, were collected when the process was thought to be in-control. The traditional Shewhart $\bar{X}$ and R charts provide no indication of an out-of-control condition, so these "trial" limits were adopted for use in on-line process control.

For the proposed MW chart with $m = 125$, $n = 5$ and $ARL_0 = 400$, we find the upper control limit $U_{mn} = 540$ and hence the lower control limit $L_{mn} = 85$. Having found the control limits, prospective process monitoring in phase II begins.

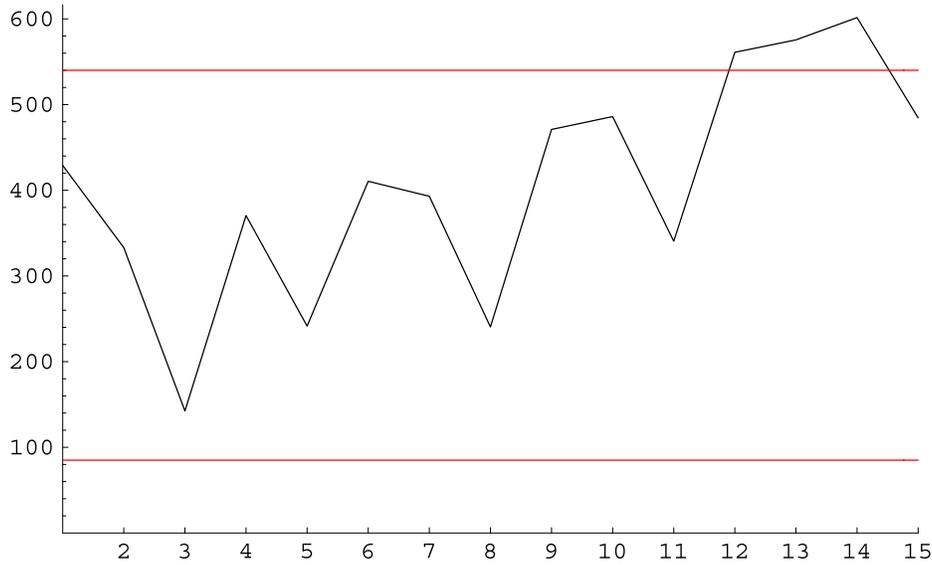

Fig 1. *MW Chart for the Piston-ring data.*

Montgomery also gives (Table 5.2) fifteen additional samples from the piston-ring manufacturing process.

These "test samples" lead to fifteen MW statistics calculated using Minitab: 429.0, 333.0, 142.5, 370.5, 241.5, 410.5, 393.0, 240.5, 471.0, 486.0, 340.5, 561.0, 575.5, 601.5 and 484.5. Comparing each statistic with the control limits, all but three of the test groups, 12, 13 and 14 are declared to be in-control. The control chart is shown in Figure 1.

The conclusion from this chart is that the medians of test groups 12, 13 and 14 have shifted to the right in comparison with the median of the in-control distribution, assuming that $G$ is a location shift of $F$. It may be noted that the Shewhart $\bar{X}$ chart shown in Montgomery [11] led to the same conclusion with respect to the means. Of course, the advantage with the MW chart is that it is distribution-free, so that regardless of the underlying distribution, the in-control $ARL$ of the chart is roughly equal to 400 and there is no need to worry about (non-) normality, as one must for the $\bar{X}$ chart. To see how the MW chart compares with other available nonparametric charts, we calculated the distribution-free precedence chart (see CVV) for this data. We found $LCL = 73.982$ and $UCL = 74.017$ for the precedence chart, with an attained $ARL_0 \approx 414.0$. Consequently, the precedence chart declares the $12^{th}$ and the $14^{th}$ groups to be out of control but not the $13^{th}$ group, unlike both the MW and the Shewhart chart.

*Comparison with the Shewhart chart*

The performance of a control chart is usually assessed in terms of its run length distribution and certain associated characteristics, such as the $ARL$. While following the general norm in the literature we examine the $ARL$, given the skewed nature of the run length distribution, we also consider two other criteria for evaluating and comparing the performance of the MW chart and its parametric competitor, the Shewhart $\bar{X}$ chart, in terms of some percentiles of the conditional run length dis-tribution. We believe these criteria provide additional useful information regarding chart performance, with estimated parameters.

To ensure a fair comparison between the MW and the Shewhart $\bar{X}$ chart, first, the $\bar{X}$ chart is used for the case when both the mean and the variance are unknown, with parameters estimated from the reference sample. Second, the charts are both designed to have the same specified $ARL_0$. Note that for the $\bar{X}$ chart the non-robustness of the $ARL_0$ with respect to non-normal in-control distributions is a major concern. This has been recognized as a problem elsewhere (see e.g. CVV) and is perhaps one of the important reasons for considering a nonparametric chart in practice.

Although both the conditional and the unconditional distributions provide important information regarding the performance of a control chart, we argue that the conditional distribution might be preferred from a practical point of view since the unconditional distribution is what results after "averaging" over all possible reference samples. Users would most likely not have the benefit of averaging in a particular application. Also, since the distribution is skewed, the percentiles and not the average are better measures of performance and thus the standard deviation appears to be a less suitable measure of variability. First we consider the in-control case.

## 2.7. In-control performance

For the in-control situation, a lower order percentile (say the $5^{th}$) is more useful and relatively large values of this percentile are desirable (in the same spirit that the $ARL_0$ of a chart be large), since that would lead to a smaller probability of a false alarm. We also show the estimated standard deviations to give an indication of the variability of $ARL_0(X)$, since this appears to be the current norm. For completeness, the $95^{th}$ percentile is also given. The two percentiles can also provide an indication of the variability in the conditional distribution.

The proposed MW chart for location for is compared to the Shewhart $\bar{X}$ chart with estimated parameters. There may be some interest in comparing against other control charts and we comment on this aspect later. To ensure a fair comparison, chart constants were determined such that the $ARL_0$ approximately equals 500 for both charts. We kept the test sample size constant, $n = 5$, and used several values for the reference sample size $m$. Both samples were drawn from a Normal(0,1) distribution. The number of simulations, $K$, was set to 1000. The results are shown in Table 4.

To illustrate, suppose $m = 750$. While applying the MW chart with $U_{mn} = 3258$, from Table 4 we know that 95% of the in-control $ARL$'s (for a large number of reference samples taken from the in-control process) will be at least 360. This provides useful performance information in addition to saying that the unconditional $ARL_0$ is 500. For the $\bar{X}$ chart with $m = 750$ and a control chart constant of 3.089 (this guarantees $ARL_0 = 500$ when parameters are both unknown), the $5^{th}$ percentile is 314 so that 95% of the in-control $ARL$'s are at least 314. Since the in-control $5^{th}$ percentiles for the MW chart are considerably larger than those of the Shewhart chart for all $m$, we conclude that the in-control performance of the MW chart is superior to that of the Shewhart chart with estimated limits, particularly for $m \leq 150$. Thus the MW chart is more useful in applications where a large amount of reference data might not be available. The uniformly smaller standard deviations for the MW chart doubly confirm its superiority. Note also that as $m$ increases, both the $5^{th}$ as the $95^{th}$ percentiles approach the mean of the corresponding unconditional distribution, the $ARL_0$, which is set at 500.

TABLE 4
*Fifth and ninety-fifth percentiles and standard deviations of the conditional in-control distribution of $ARL_0(X)$; All cases: $n = 5$ and $ARL_0 = 500$*

| $m$ | Upper control limit MW | $5^{th}$ perc. MW | $95^{th}$ perc. MW | St. Dev. MW | Upper control limit Shewhart | $5^{th}$ perc. Shewhart | $95^{th}$ perc. Shewhart | St. Dev. Shewhart |
|---|---|---|---|---|---|---|---|---|
| 50 | 217 | 97 | 1292 | 553 | 3.01996 | 49 | 1619 | 854 |
| 75 | 326 | 146 | 1219 | 461 | 3.05156 | 87 | 1379 | 645 |
| 100 | 435 | 182 | 1146 | 358 | 3.06535 | 112 | 1290 | 463 |
| 150 | 654 | 251 | 1090 | 315 | 3.07715 | 154 | 1197 | 377 |
| 300 | 1304 | 284 | 845 | 197 | 3.08607 | 232 | 927 | 235 |
| 500 | 2172 | 322 | 700 | 140 | 3.08848 | 270 | 828 | 174 |
| 750 | 3258 | 360 | 677 | 107 | 3.08935 | 314 | 765 | 140 |
| 1000 | 4347 | 379 | 674 | 83 | 3.08969 | 338 | 721 | 121 |
| 1500 | 6520 | 409 | 642 | 71 | 3.08996 | 367 | 678 | 97 |
| 2000 | 8691 | 420 | 629 | 55 | 3.09007 | 376 | 651 | 84 |

### 2.8. Out-of-control performance

The distribution-free property (and the resulting robustness of the $ARL_0$) is an important asset of the proposed chart, but what about its out-of-control performance? We address this issue here. Since this is a chart for location, our interest is in the shift (location) alternative $G(x) = F(x - \delta)$, where $\delta$ is the unknown shift parameter. To study the effects of using a reference sample, again, we use conditioning and study the distribution of the conditional $ARL$. Let $ARL_\delta(X) = E_{F(x-\delta)}(N|X)$, denote $ARL$ of the run-length distribution, given the reference sample $X$, when the process distribution $F$ has shifted by an amount $\delta$. We examine the out-of-control performance of both the MW chart and the Shewhart chart in terms of $ARL_\delta(X)$ next.

We also provide a more traditional chart comparison by examining the out-of-control unconditional $ARL$ for a specified distribution and shift, $ARL_\delta$. Naturally, small values of $ARL_\delta$ are desirable. As in the case of the in-control situation, we also examine a percentile of the out-of-control distribution. However, in the out-of-control case it makes sense to focus on a higher order percentile, say the $95^{th}$ percentile. Denoting this by $q_{0.95}$, relatively smaller values of $q_{0.95}$ are desirable for a preferred chart, since the probability of a signal is desired to be higher in the out-of-control case. For a given value of $q_{0.95}$, users can be 95% confident that for their own specific reference sample, the out-of-control $ARL$ is smaller than that value. The two performance measures, namely the $ARL_\delta$ and the $q_{0.95}$, are examined for three distributions: Normal, Laplace and Gamma(2,2). The Laplace distribution is normal-like but with heavier tails, which results in higher probabilities of extreme values. The Gamma(2,2) distribution is skewed and is often used in the SPC literature. We apply two-sided charts to the Normal and the Laplace distributions and a one-sided chart with an upper control limit to the Gamma(2,2) distribution. The test sample size $n$ is 5 and the reference sample size $m$ is 100. Control limits for both the MW chart and the Shewhart $\bar{X}$ chart with estimated parameters are determined such that $ARL_0 = 500$. Using these limits $ARL_\delta(X)$ and the $95^{th}$ percentile of the distribution of $ARL_\delta(X)$ are computed for several values of $\delta$, which is given in units of the standard deviation. Figures 2 through 4 show the results.

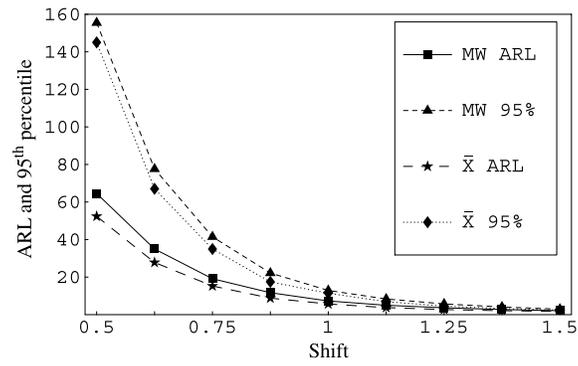

FIG 2. *Performance for MW chart and Shewhart chart under Normal shift alternatives.*

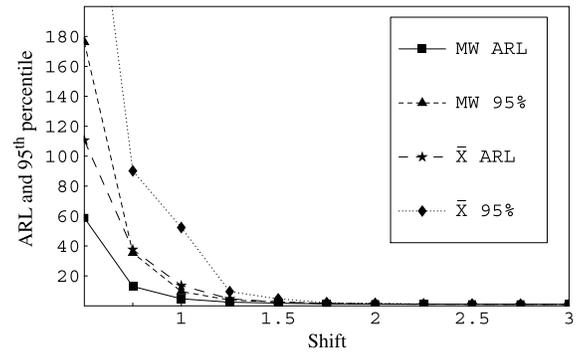

FIG 3. *Performance for MW chart and Shewhart chart under Laplace shift alternatives.*

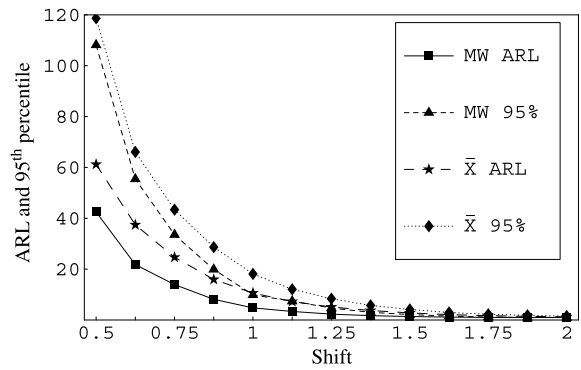

FIG 4. *Performance for MW chart and Shewhart chart under Gamma(2,2) shift alternatives.*

For the set of points, triangles and diamonds, observe that for the normal distribution the $95^{th}$ percentiles for the Shewhart $\bar{X}$ chart are all smaller than those for the MW chart. Thus, as one might expect, the $\bar{X}$ chart is more effective in detecting shifts than the MW chart in case of the normal alternative. However, note that the differences between the percentiles are small at all shifts (the largest difference is around 15) and the difference appears to vanish for shifts greater than 1. The same pattern holds for the two $ARL$'s. On the other hand, Figure 3, for the Laplace distribution, shows that the MW chart is clearly better than the Shewhart chart for all shifts, large and small. For the Gamma(2,2) distribution, in Figure 4, again, we see that the MW chart is better in detecting shifts, although the difference in performance is not as dramatic as in the case of the Laplace distribution. These calculations were repeated for $m = 500$ observations; the results were very similar and are therefore omitted here. We conclude that the nonparametric MW chart follows the well-known results for the MW test statistic: it is nearly as effective as the Shewhart $\bar{X}$ chart under normality, but is more effective under heavy tailed and skewed distributions. Also, note that performance of the MW chart in the case of the Laplace distribution makes it potentially useful when outliers in the data are not uncommon.

## 3. Discussions and further topics

*Comparison with CUSUM and EWMA charts*

Shewhart charts are known to be very good for moderate to large shifts. These charts do not require tuning parameters like those needed by the CUSUM and the EWMA charts for a specified shift, but aim for global performance. Thus although one can design a CUSUM or an EWMA chart to perform better by focusing on either small, medium or large shifts, a priori, the same path could lead to a worse performance for other shifts. Moreover, as has been noted in the literature Quesenberry [14], the very nature and principle of the "averages-type" charts (such as CUSUM and EWMA) is different from that of Shewhart charts. In addition to the problem of having different in-control run-length distributions that renders the stable properties of these charts to be quite different (and hence out-of-control assessments less meaningful), the averages charts are more powerful in detecting specific types of shifts (sustained monotonic).

Nevertheless, we made an attempt to compare the MW chart with EWMA and CUSUM on the basis of in-control robustness in terms of misspecifications of the shape of the distribution and the variance. Rather than showing all the results, we summarize the findings here. For fast detection of large shifts (say 1 and larger), the MW chart is very good: it is powerful and maximally robust against misspecifications of shape and variance. The latter is far from being true for CUSUM and EWMA designed for detecting large shifts: true $ARL_0$ could easily be twice as large or small as the target $ARL_0$ in case of skewness, increased variance or heavy-tails. For small shifts, the situation is more delicate: the EWMA is a strong competitor, since it is often (but not always; for example, not for medium to large shifts in the normal) more powerful than the MW chart and quite robust against misspecifications of the shape. This type of robustness of the EWMA chart was shown earlier (see e.g. Borror et al. [2]). However, the EWMA is seen to be not robust against a misspecification of the variance. For example, a small increase of the variance of a normal (from 1 to 1.1) lowered the in-control ARL from 500 to 215 for an EWMA designed for small shifts. The overall conclusion is that especially when the in-control variance can be estimated with limited accuracy, the nonparametric charts in general, and the MW chart in particular, is extremely useful in practice, because they do not require knowledge of the underlying variance. Similar conclusions have been drawn in Amin et al. [1].

*Alternative chart design criteria*

Table 4 suggests that with estimated parameters, especially for small values of $m$, it may be useful to use a lower order (say the $5^{th}$) percentile of the conditional distribution as a chart design criterion rather than the mean, i.e., the unconditional mean $ARL_0$. The design requirement would be that the percentile be at least equal to some specified large number, such as 300. In some cases one might want to avoid very short in-control runs in the future, which suggests using a lower percentile of the in-control distribution of $N$ and not that of the conditional distribution of $E(N|X)$. The probability $P_0(N \leq n)$ can be computed using similar methods as for computation of $ARL_0$. Then, for example, if one wants to avoid in-control runs smaller than 100 with a high probability, say 0.90, we can find the control chart limits by solving $P_0(N \leq 100) = 1 - 0.90 = 0.10$, again using a search algorithm. To facilitate use of both of these chart design criteria we have implemented these in the software that we provide with this paper.

*Individual's chart*

There is considerable interest in nonparametric charts for individual observations ($n = 1$) since in this case the notion of approximate normality via the central limit theorem is not applicable. For $n = 1$, formula (2.7) allows for fast exact computations. The software can be used to set up such a chart. Because of the natural interest in nonparametric individual's charts, a detailed treatment of this topic will be given in a future paper.

*Monitoring dispersion*

The dispersion or spread need not be monitored while using a nonparametric control chart (such as the MW chart) under the location model. This has been cited as an advantage of the nonparametric sign charts by some authors. However, monitoring the spread is an interesting practical problem in a "location-scale" model and we see the possibility of designing a chart for scale (along with that for the location) based on some nonparametric test for scale, This topic will be considered in a future paper.

**Appendix: Software**

In order to support practical implementation of the methods presented in this paper two types of software related resources are provided. First, a Mathematica 4.2 Wolfram [16] notebook is available to calculate (i) the in-control ARL computations any of the five methods (ii) the out-of-control performance computations, (iii) the control chart constants and (iv) to plot the MW-control chart for a user-specified data set. Second, we created a website that enables anyone to apply the proposed methodology. The MW control chart limits can be found for the sample sizes at hand for a specified target $ARL_0$ (or a desired ($q$th) percentile of $ARL_0(X)$). Moreover, the website allows users to import their own data set and have the MW chart drawn. The site can be reached via [www.win.tue.nl/∼markvdw](www.win.tue.nl/∼markvdw). The Mathematica notebook is available from the same site; it contains more procedures and allows for more flexible input. User instructions are available in the notebook and at the website.

**Acknowledgments.** We thank Marko Boon (TU Eindhoven) for assistance with the design of the Mann-Whitney control chart website.